# KoSpeech: Open-Source Toolkit for End-to-End Korean Speech Recognition

***Soohwan Kim[1], Seyoung Bae[2], Cheolhwang Won[3], Suwon Park[4]***

Kwangwoon University

[1]kaki.brain@kakaobrain.com, [2,3]{triplet02, wch18735} @naver.com, [4]spark@kw.ac.kr

## ABSTRACT

***KoSpeech***, an open-source software, is modular and extensible end-to-end Korean automatic speech recognition (ASR) toolkit based on the deep learning library PyTorch. Several automatic speech recognition open-source toolkits have been released, but all of them deal with non-Korean languages, such as English (e.g. ESPnet, Espresso). Although AI Hub opened 1,000 hours of Korean speech corpus known as KsponSpeech[1], there is no established preprocessing method and baseline model to compare model performances. Therefore, we propose preprocessing methods for KsponSpeech corpus and a baseline model for benchmarks. Our baseline model is based on Listen, Attend and Spell (LAS) architecture and ables to customize various training hyperparameters conveniently. By *KoSpeech*, we hope this could be a guideline for those who research Korean speech recognition. Our baseline model achieved 10.31% character error rate (CER) at KsponSpeech corpus only with the acoustic model. Our source code is available here[2].

## 1. INTRODUCTION

End-to-end automatic speech recognition is an emerging paradigm in the field of neural network-based speech recognition that offers multiple benefits. Traditional hybrid ASR systems consist of complicated structures, in which a DNN-HMM-based acoustic model [1, 2, 3, 4, 5, 6], vocabulary dictionary, and language model are used to construct a single decoding network. To reduce this complexity, an end-to-end ASR system handles all this process with a single system. Because of this convenience, many End-to-End models have been proposed such as Listen, Attend and Spell (LAS) [7]. LAS is based on the sequence to sequence learning framework with attention [8, 9, 10, 11, 12]. It consists of an encoder recurrent neural network (RNN), which is named *listener*, and a decoder RNN, which is named *speller*. LAS is simple in structure, intuitive, and accessible even without domain knowledge.

[1]http://www.aihub.or.kr/aidata/105

[2]https://github.com/sooftware/KoSpeech/

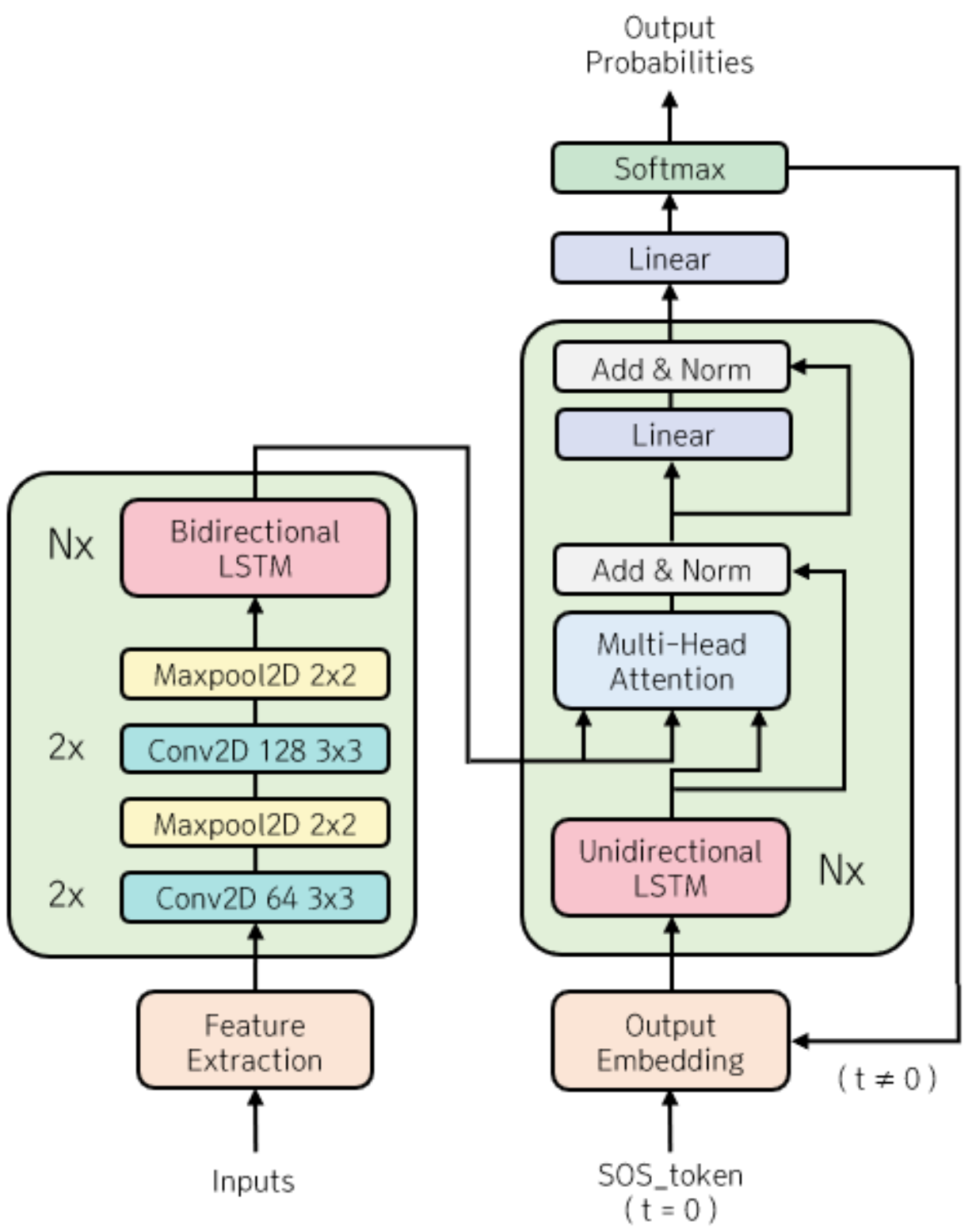

Figure 1: ***Baseline architecture.*** *Our baseline model consists of two parts: encoder RNN with deep CNN (listener), decoder RNN with attention mechanism (speller).*

Because of these advantages, many variant acoustic models proposed are based on LAS have been proposed [13, 14, 15], and there were many improvements in performance. Thanks to this easier access to speech recognition, currently several speech recognition implementations have been released [16, 17]. However, most open sources are limited to English corpus such as LibriSpeech, WSJ, Switchboard, CallHome [18, 19, 20, 21]. This has become a major factor in raising entry barriers to Korean speech recognition.

In 2018, AI Hub released a large amount of Korean dialog dataset, KsponSpeech. KsponSpeech consists of 1,000 hours of Korean dialog of various domains, which 2,000 different speakers have recorded in quiet condition. Although a couple of years have passed since the KsponSpeech corpus was released by AI Hub, it is hard to compare model performance with each other because there are no preprocessing methods firmly established, and no baseline model is provided. For these reasons, we introduce *KoSpeech*, an open-source framework for end-to-end Korean speech recognition to make a guideline for further works with KsponSpeech corpus, and therefore comparing performances with each other. In addition to providing codes for the core speech recognition tasks, KoSpeech was designed by the principle of triple E (EEE): ($E_1$) Easy to use, ($E_2$) Easy to read, ($E_3$) Easy to expand.

We also dealt with pre-processing methods for KsponSpeech. There are several issues concerning preprocessing KsponSpeech corpus, such as selecting between phonetic transcription and spelling transcription selection, special character removal, etc. We hope our research can be a guideline for later works. Also, there was an attempt to perform Korean speech recognition on a grapheme unit basis [22]. To help these researchers, we have configured it to preprocess in character, subword, and grapheme unit.

On the model part, we have organized various structures based on LAS. We employ two convolution neural networks (CNN) feature extractor proposed in [23, 24]: (Deep Speech 2 and VGG Net), and four attention mechanisms proposed in [10. 12. 25, 26]: (Dot-Product, Additive, Multi-Head and Location-Aware). Our baseline model consists of VGG extractor and a Multi-Head attention mechanism. Other details will be covered in the model section of Chapter 3.

This technical report describes the technologies used in the KoSpeech, starting with speech signals, how the acoustic model was constructed, data augmentation, etc. We end by showing the results of the system`s benchmarks and the experiments conducted in terms of accuracy in KsponSpeech.

## 2. RELATED WORKS

Sequence to sequence framework proposed to solve the problems with learning and predicting input and output sequences of variable-length [8]. Encoder RNN is used to make fixed-length vectors out of the variable-length input. Then decoder RNN converts this vector to produce the variable length output sequence back, one token at a time.

This sequence to sequence framework has been widely used for many applications such as machine translation [27, 28], image captioning [29, 30], parsing [31] and conversational modeling [32]. Speech recognition can also be a direct application [11, 12], because of the generality of the framework.

End-to-end speech recognition is an active area of research. Many research and papers like [7, 22] are announced, and multiple ways to improve efficiency and accuracy are proposed. One of them is an attention mechanism, that provides decoder RNN more information when it produces the output tokens [10]. Adding an attentional mechanism to the decoder shows a dramatic improvement of the model performance, particularly with long inputs or outputs [10]. In speech recognition tasks, the RNN encoder-decoder with attention performs well both in predicting phonemes [11] or graphemes [7, 33].

Furthermore, large scale public speech corpora allowed many ASR models to be able to handle many real-world applications. Early days, which is in the 1990s, public speech corpora were released such as LibriSpeech, Wall Street Journal, Switchboard, CallHome [18, 19, 20, 21]. These datasets are still influential as benchmark datasets these days [34, 35, 36]. Recently, however, Librispeech [18] is the most popular benchmark speech corpus on which the latest state-of-the-art ASR models are evaluated [22, 37, 38]. Although they are greatly useful and dependable, existing speech corpora mainly deal with English or non-Korean language.

In 2018, AI Hub released 1,000 hours of Korean dialog dataset, KsponSpeech. This corpus consists of 1,000 hours of Korean dialog of various domains, which 2,000 different speakers have recorded in the quiet condition. Thanks to the KsponSpeech, entry barriers of Korean Speech recognition has been lowered a lot.

## 3. ASR SYSTEM

In this section, we address the ASR system, in detail, with transcript pre-processing, speech signals, and acoustic models and optimization. We tried to write as much detail as possible for those who are new to Korean speech recognition.

### 3.1. Transcript Preprocessing

KsponSpeech dialog transcript has followed ETRI transcription rules, so there are numerous unnecessary tokens for ASR tasks. Through a series of methods below, original scripts can be converted into appropriate scripts for ASR tasks.

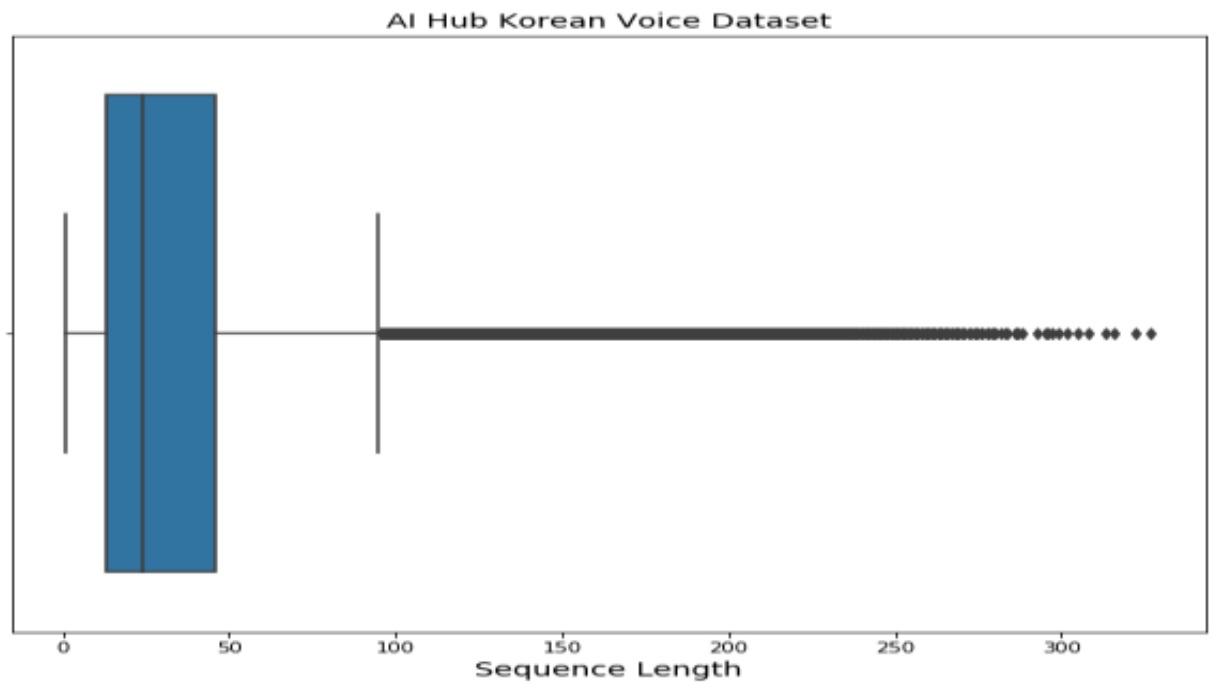

Figure 2: ***Boxplot of KsponSpeech script file length.*** *At original data, files that have sequence length over 100 have been classified as an outlier.*

#### *3.1.1. Statistical Analysis*

There are more than 620,000 text files in the original KsponSpeech corpus. As in Figure 2, most of the script has a sequence length of less than 100, calculated via character-unit based. We construct training files which have sequence length longer than 100 to make the model to train faster, and to lessen memory usage. Moreover, it was also because lengthy sentences can act like as noise while training since conversations are not that much longer in actual ASR situations. By this method it was able to improve training speed a little and reduce memory usage significantly since extremely long sentences compared to other sentences in one batch are not used.

### 3.1.2. Special Token Handling

According to ETRI transcription rules, there are some special tokens to indicate background noise, speaker's breathing sound, etc. Furthermore, transcription also indicates meaningless interjections. We deleted those tokens because we thought they were unnecessary for ASR tasks.

Table 1*: **Example of preprocessing sequence.** The original script from KsponSpeech (upper), a preprocessed script that deleted special tokens (middle), and a script that also deleted interjection tokens (lower).*

| | Transcript |
|---|---|
| # original | b/ 아 (70%)/(칠 십 퍼센트) 확률이라니 모+ 몬 소리야 n/ |
| # select phonetic | 아 칠 십 퍼센트 확률이라니 모+ 몬 소리야 n/ |
| # delete noise label | 아 칠 십 퍼센트 확률이라니 모+ 몬 소리야 |
| # delete interjection | 아 칠 십 퍼센트 확률이라니 모 몬 소리야 |

### 3.1.3. Selective Transcription

In the original script, there are original, or grammatically correct, expressions at first parenthesis, and phonetic expressions, even if the speaker has pronounced incorrectly, at second parenthesis.

Table 2: ***Examples of phonetic (upper) and original, or spelling (middle) transcriptions.** The Last option is to choose the original transcriptions only if it is a numeric expression.*

| | Transcript |
|---|---|
| # phonetic transcript | 아 칠 십 퍼센트 확률이라니 모 몬 소리야 |
| # spelling transcript | 아 70% 확률이라니 뭐 뭔 소리야 |
| # hybrid transcript | 아 칠 십 퍼센트 확률이라니 뭐 뭔 소리야 |

By KoSpeech, this can be selectively accepted via options. We generally accept phonetic expressions, even if it is ungrammatical because it is written as it sounds, acoustic data will be much more similar to that expression. However, in some cases like numeric expressions, researchers can determine which one to choose, by setting options. See table 2.

### 3.1.4. Variable Base Unit

KoSpeech provides processing in characters, subwords, and grapheme units as visualized in table 3. In this paper, we are considering character-unit based conditions only.

Table 3: ***Example of a script processed by various base units.** Subwords are segmented using SKT`s pre-trained KoBERT tokenizer or Google`s sentencepiece.*

| | Transcript |
|---|---|
| # character unit | 아 모 몬 소리야 |
| # subword unit | _아 _모 _ 몬 _소리 야 |
| # grapheme unit | ㅇㅏ ㅁㅗ ㅁㅗㄴ ㅅㅗㄹㅣㅇㅑ |

To tokenize into subwords, we employed sentencepiece, or SKT`s pre-trained KoBERT tokenizer (optional) [39, 40]. Plus, we use hgtk (hangul-toolkit) for the graphene unit tokenizer [41].

## 3.2. Speech signals

KsponSpeech audio files have a format of 16KHz/16bits of sample/bit rate, headerless (little-endian) linear pulse-code modulation (PCM). Through a series of methods below, original audio can be trimmed or modified for faster training and better model performance.

### 3.2.1. Silence Elimination

There are quite a few silence sections in original KsponSpeech audio files.

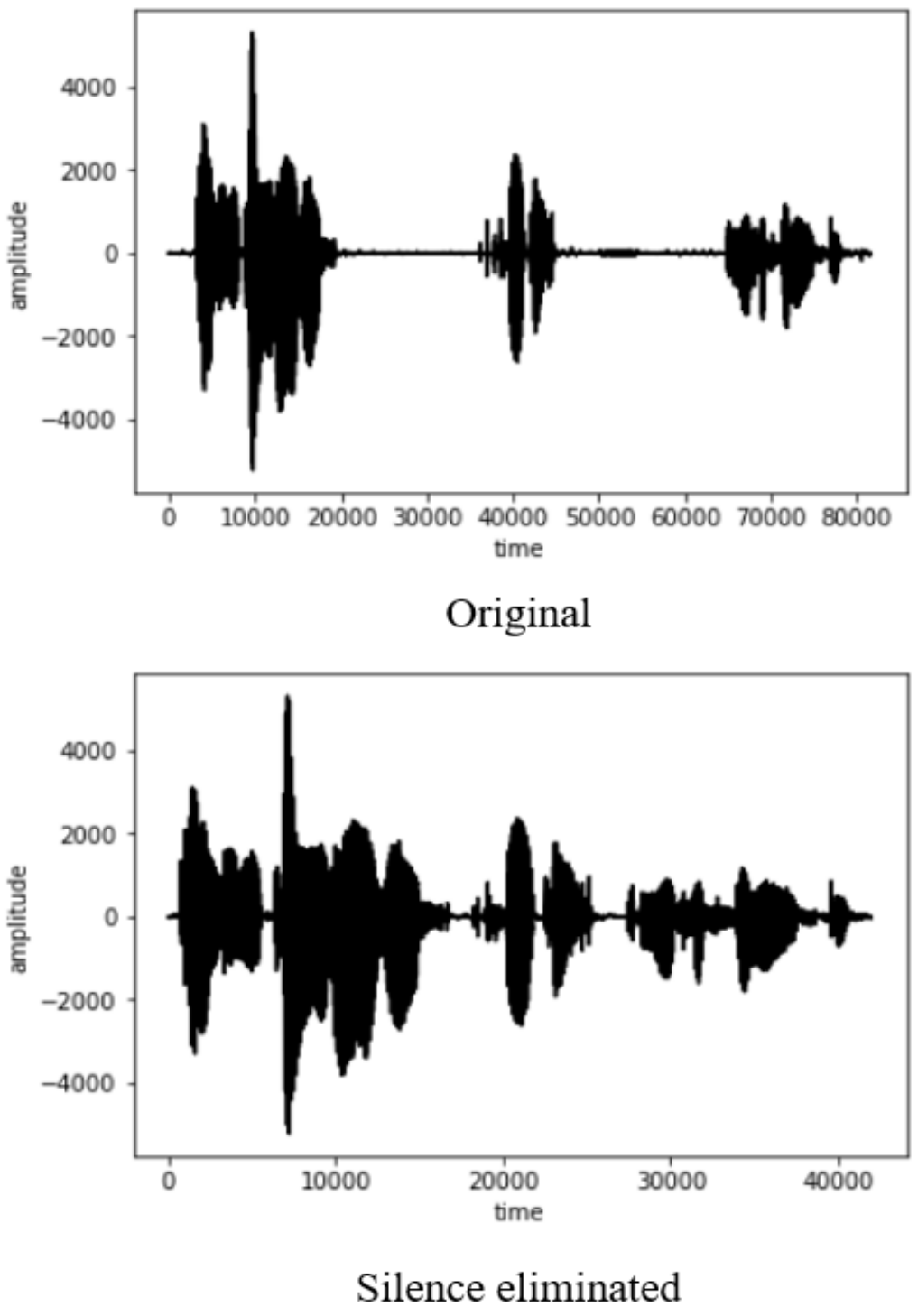

Figure 3: ***Example of two silence sections in the original audio file eliminated by threshold decibel (30dB).** The Total length reduced to half, and conversation was understandable before and after silence elimination.*

This section contains no information of the conversation, so we simply eliminated these silence sections by threshold of 30db. This threshold decibel has been chosen by multiple experiments. We confirmed there was no loss of verbal content before and after elimination at all test cases. Since length of the audio file can be reduced significantly although there is a risk of a little loss of the conversational information, the model can train faster than using original audio untrimmed.

*3.2.2. Feature Extraction*

We used various features like Mel-Frequency cepstral coefficients (MFCCs), log mel spectrogram, log spectrogram, filterbank. In the extracting process, parameters like frame length and stride, windows effects on performances. The process by which these values are calculated will be described below.

**Windowing** The input speech signal is segmented into frames of 10 ~ 25ms with overlap. Usually the frame size is equal to power of the two in order to facilitate the use of fast fourier transform (FFT). If this is not the case, the rest of the samples are filled with zero to the nearest length of power of two. In this case, cutting digital data into each segment inevitably makes a discontinuous point which causes a high frequency in the spectrum. Therefore, each frame is multiplied with a hamming window in order to keep the continuity of the first and the last points in the frame.

$$\mathrm{w(n)} = 0.54 - 0.45 * \cos\left(\frac{2\pi n}{N-1}\right) \text{ where } 0 \leq n \leq N-1$$

**Spectrogram** To sum up briefly, spectrograms are like a bunch of FFTs stacked on top of each other. A spectrogram is a visual representation of the spectrum of frequencies of a signal as it varies with time. Spectrograms are used not only in the field of speech recognition but also of music, speech synthesis, speech classification, etc. Spectrogram has various information types but in speech recognition, we mainly focus on the formant which corresponds to a resonance in the vocal tract. A formant by which our auditory organ discriminates pronunciation in speech is concentration of acoustic energy around a particular frequency in the speech wave. Each 10~25ms audio segment has their own spectrogram and they provide the formant information and envelope.

Spectrogram is calculated from the time signal using the fourier transform, especially Fast Fourier Transform. Digitally sampled data in time domain is segmented into chunks. Usually appropriate segmentation length was 20~25ms in our experiment. Also, appropriate overlapping proportion was 25% or 50%. Each segment is fourier transformed to calculate the magnitude of the frequency spectrum. Each result is then laid side by side to form the spectrogram Figure 7. This process essentially corresponds to computing the squared magnitude of the short time fourier transform (STFT) of signal. It is represented as following relation with window width $w_i$.

$$Spectrogram(t, w) = |STFT(t, w)|^2$$

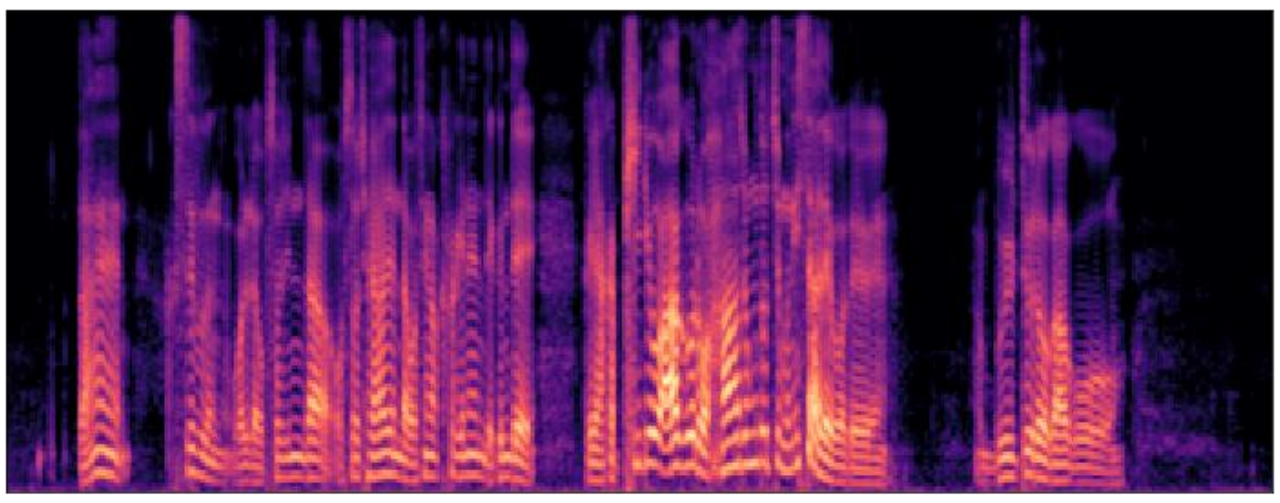

Figure 4: ***Mel spectrogram image.*** *Vertical axis shows spectrum of segmented sample and each sample is aligned along time axis.*

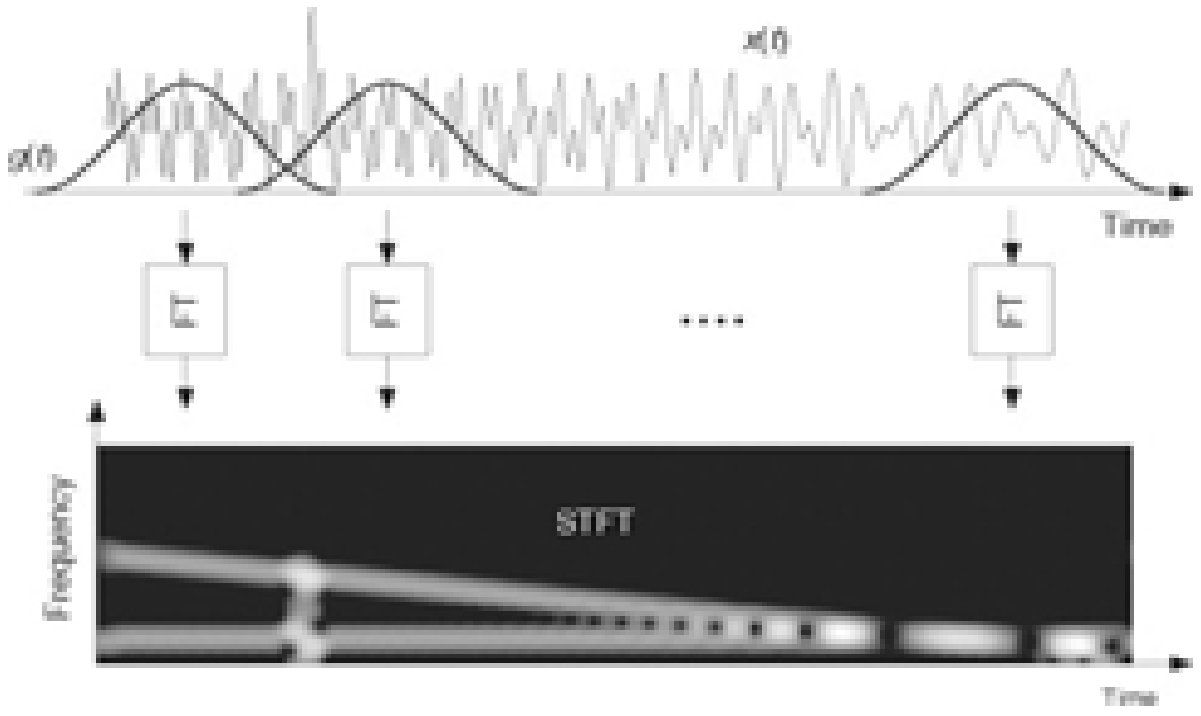

Figure 5: ***Spectrogram summarized to image.***

**Mel spectrogram** Humans do not perceive frequencies on a linear scale. We are better at detecting differences in lower frequencies than higher frequencies. For example, we can easily tell the difference between 500 and 1,000 Hz but will hardly be able to tell the difference between 10,000 and 10,500 Hz, even though the distance between the two pairs are the same. This unbalanced perception phenomenon is caused by the auditory organ. Each different frequencies stimulate corresponding basilar membrane and the critical bandwidth, any range of frequencies within which any two signals strongly stimulate a common portion of the basilar membrane, become larger proportionally to a logarithm of a frequency. Stevens, Volkmann, and Newmann proposed a unit of pitch such that equal distances in pitch sounded equally distant to the listener. This is called the mel scale.

There is no single mel-scale formula, so we introduce the popular formula from O'shaughnessy's book.

$$\mathrm{m} = 2595 \log_{10}\left(1 + \frac{f}{700}\right)$$

The main difference between spectrogram and mel spectrogram is a frequency domain scale. A mel spectrogram is a spectrogram where the frequencies are converted to the mel scale above, so we can take a form similar to the level humans actually hear at.

**Log scaled spectrogram** So far, we have looked into the spectrogram and mel spectrogram. At the end, we are able to apply auditory theory to each calculated power. For example, humans more easily notice the intensity gap in lower power levels than higher power levels. For that reason, log spectrogram, log mel spectrogram are calculated by logarithm.

**MFCC and Mel filter bank** Last feature is mel-frequency cepstral coefficient (MFCC). Before we find out how to calculate the MFCC, we need to know what cepstrum is. Acoustic model consists of excitation and formant.

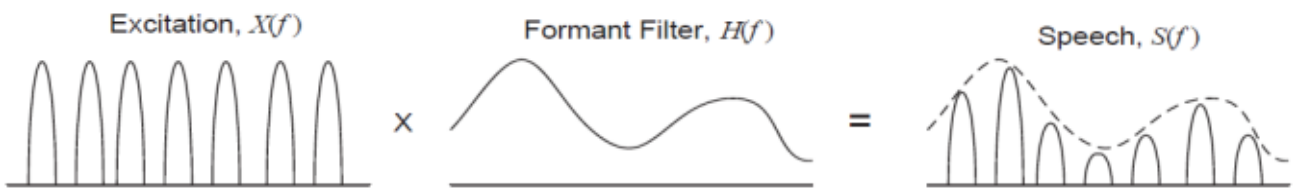

Figure 6: ***Speech signal in frequency domain***

Our goal is to get a spectral envelope H(f) slowly changing spectrum from S(f). It is a low time component. So, to filter high time components, excitation spectrum X(f), apply log operation to change multiplication into summation.

$$log|S(f)| = log|X(f)| + log|H(f)|$$

Finally, by inverse fourier transform, we can get a low-time spectrum. This spectrum is called ceps-trum which is the reverse order of spec. Also, the process to filter high time components is called as liftering which is reverse order of fil.

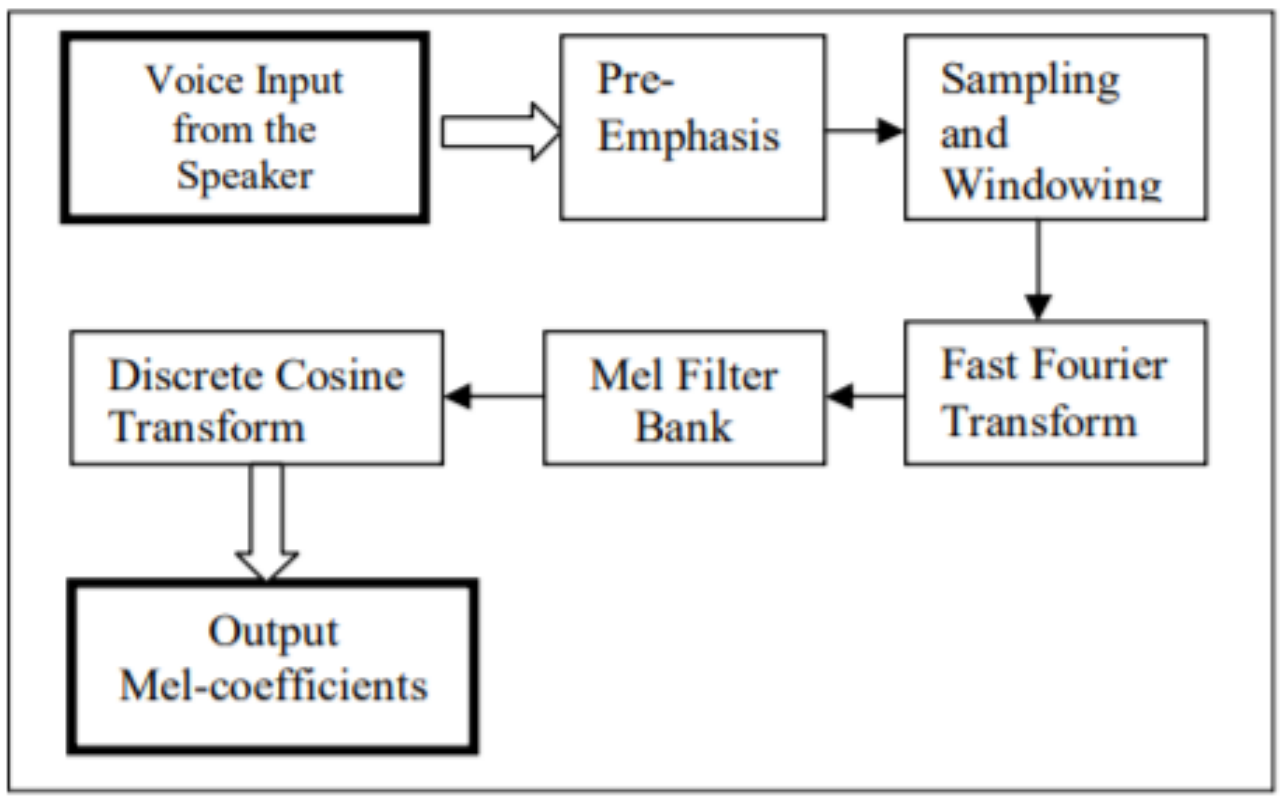

Figure 7: ***Block diagram for MFCC***

MFCC follows a similar process. In sound processing, the mel-frequency cepstrum (MFC) is a representation of the short-term power spectrum of a sound, based on a linear cosine transform of a log power spectrum on a nonlinear mel scale of frequency. This MFCCs algorithm is generally preferred as a feature extraction technique to perform voice/speech recognition. The step-by-step computation of MFCC is explained in following Figure 10 from [42].

The magnitude frequency response is multiplied by a set of triangular band pass filters to get the log energy of each triangular band pass filter. The positions of these filters are equally spaced along the mel frequency and weighted sum of filter and $|S(f)|$ is called $fbank_i$, mel-band energy is also called the mel filter bank feature.

This mel-band energy goes through non-linear transformation $f_i = \ln(\text{fban}k_i)$. If you choose the number of the filters to 23, discrete cosine transform of $f_i$ would take only $0 \leq i \leq 12$ for liftering high time components.

$$C_i = \sum_{j=1}^{23} f_i \cos\left(\frac{i \times \pi}{23}(j - 0.5)\right), \; 0 \leq \text{i} \leq 12$$

After following discrete cosine transform, final output would be 13 MFCC and log energy.

$$\text{logE} = \ln(\textstyle\sum_{i=0}^{N-1} \text{s[i]}) \text{ where } s[i] \text{is a speech signal.}$$

**Summary** Transformation spectrogram to mel-spectrogram or mel frequency cepstral coefficient (MFCC) is kind of dimensional reduction. These methods are very popular and show high performance before the appearance of a speech recognition model based deep neural network. Because of expectation from high model power, we also adopt the spectrogram. Log scaled spectrogram and mel spectrogram are transformed based on auditory theory and normalization for medel training.

### 3.2.3 SpecAugment

Data augmentation has been proposed as a method to generate additional training data for ASR. [43] proposes the SpecAugment integrating above methods. In the [43], They explain SpecAugment as the method that operates on the log mel spectrogram of the input audio, rather than the raw audio itself. We apply SpecAugment to not only log mel spectrogram but also spectrogram, mfcc, mel filter bank and are able to see better performance.

This method is simple and computationally cheap to apply, as it directly acts on the log mel spectrogram as if it were an image, and does not require any additional data. We are thus able to apply SpecAugment online during training. SpecAugment consists of three kinds of deformation of the spectrogram. The first is time warping, a deformation of the time-series in the time direction. The other two augmentations, inspired by “Cutout”, proposed in computer vision [44] are time and frequency masking, where we mask a block of consecutive time steps or specific band width in spectrogram. Among these, time warping method performs below our expectations, we develop time-masking and frequency-masking by ourselves.

- Frequency masking is applied so that $f$ consecutive frequency channels $[f_0, f_0 + f)$ are masked, where $f$ is first chosen from a uniform distribution from 0 to the frequency mask parameter $F$, and $f_0$ is chosen from $[0, v - f)$. $v$ is the number of channels along the frequency axis.
- Time masking is applied so that $t$ consecutive time steps $[t_0, t_0 + t)$ are masked, where $t$ is first chosen from a uniform distribution from 0 to the time mask parameter $T$, and $t_0$ is chosen from $[0, \tau - t)$.

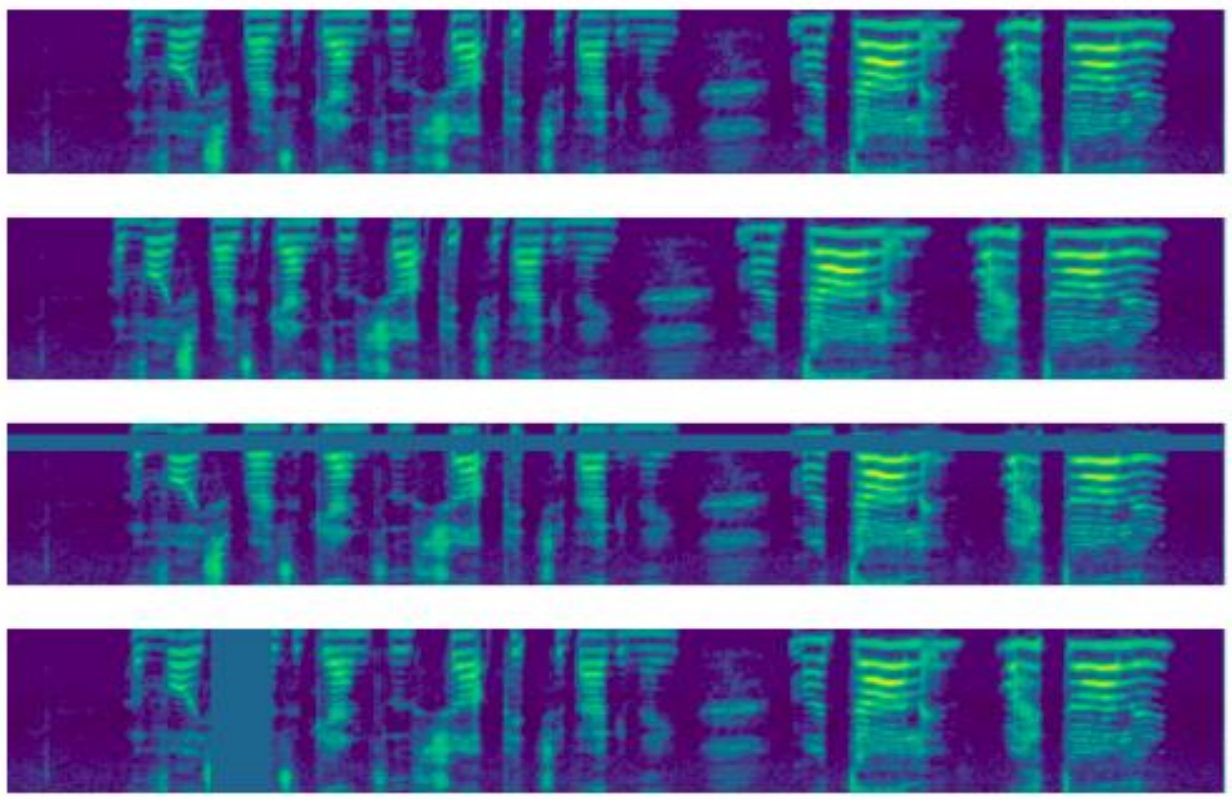

Figure 8: ***Augmentations applied to the base input, given at the top.*** *From top to bottom, the figures depict the log mel spectrogram of the base input with no augmentation, frequency masking, time masking and frequency and time masking both.*

Figure 8 shows examples of the augmentations applied to a single input.

## 3.3. Acoustic Model

We use Listen, Attend and Spell (LAS) networks [7] for our ASR tasks. (See figure 1.) Also, we employed architecture proposed in [24] except connectionist temporal classification (CTC) algorithms. These models, being end-to-end, are simple to train. In this section, we treat our baseline model and some techniques. This section addresses four topics of acoustic model: *Listener* (Section 3.3.1), *Speller* (Section 3.3.2), *Attention Mechanism* (Section 3.3.3) and *Optimization* (Section 3.3.4).

### 3.3.1 Listener

Our encoder network is boosted by using deep CNN, which is proposed in [22, 23]. The first one called Deep Speech 2 (ds2) extractor, and the second called VGG extractor. We use the initial layers of the DS2 / VGG net architecture followed by Bi-directional LSTM (BLSTM) layers in the encoder network. We used the following CNN extractor:

**Deep Speech 2 Extractor**:

Conv2dBNReLU (# in = 1, # out = 32, filter = 41 × 11)

Conv2dBNReLU (# in = 32, # out = 32, filter = 21 × 11)

**VGG Net Extractor:**

Conv2dBNReLU (# in = 1, # out = 64, filter = 3 × 3)

Conv2dBNReLU (# in = 64, # out = 64, filter = 3 × 3)

Maxpool2D (patch = 3 × 3, stride = 2 × 2)

Conv2dBNReLU (# in = 64, # out = 128, filter = 3 × 3)

Conv2dBNReLU (# in = 128, # out = 128, filter = 3 × 3)

Maxpool2D (patch = 3 × 3, stride = 2 × 2)

*Conv2dBNReLU consists of a convolution 2D layer followed by batch normalization layer and ReLU activation function.*

Input speech feature images are downsampled to images along with the time-frequency axises through the two max-pooling (Maxpool2D) layers or filters. More details of DS2, VGG extractor are described in [22, 23]. This high-level feature from CNN extractor enters input of BLSTM. The RNN module of LAS encoder consists of three stacked bi-directional LSTMs with 512 units per direction.

### 3.3.2 Speller

The decoder has two stacked unidirectional LSTM with 1024 units and two projection layers to predict the character probability distribution. The attention learns the alignment between the encoder outputs (*value*) and the decoder hidden states (*query*). Multi-Head attention is employed for the speech alignment. Multi-Head attention was proposed for Transformer [25], combination with Sequence-to-Sequence architecture is also showing good performance [13].

We also used the residual-connection technique proposed in [26] for computer vision tasks. After that, it is employed in Transformer [26] to prevent vanishing gradient problems and maintain decoder RNN`s information (query). They add directly their input into the next layer to maintain their positional encoding information after attention or feed forward network. We use this connection to maintain an upper layer gradient and make an upper layer possible to pass on gradient to lower layer without vanishing problem.

### 3.3.3 Attention Mechanism

The attention mechanism has been a fairly popular concept and a useful tool in the deep learning community in recent years. In this section, we are going to look at the attention we used. We experimented with 4 attentions [25, 10, 12, 26], and these results will be addressed in Section 4.

**Scaled Dot-Product Attention**

Scaled Dot-Product Attention consists of queries and keys of dimensions $d_k$ and values of dimension $d_v$. This scaled dot-product attention computes the dot products of the query with all keys, divides each by $\sqrt{dk}$, and applies a softmax function to obtain the weights on the values.

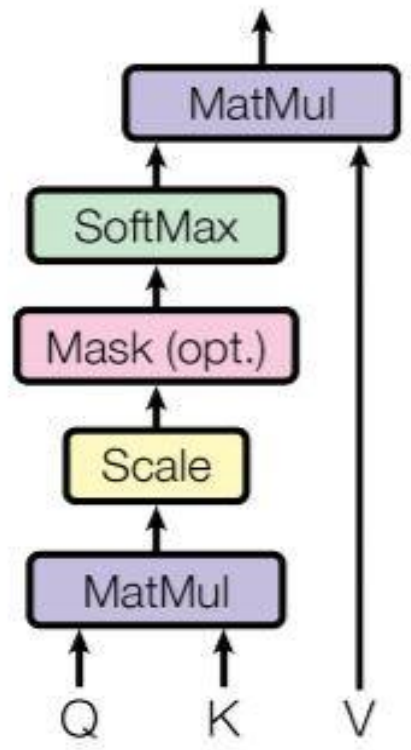

Figure 9: ***Scaled Dot-Product Attention***. *Computing dot-product attention is identical, except for dot-product of query and key is not divided by* $\sqrt{d_k}$.

This attention mechanism computes the matrix of output as:

$$Attention(Q.K.V) = softmax(\frac{QK^T}{\sqrt{d_k}})V$$

The two most commonly used dot-product (multiplicative) attention. Dot-product attention is identical to this algorithm, except for the scaling factor of $\frac{1}{\sqrt{d_k}}$. This scaling allows much faster convergence and more space-efficiency in practice.

**Additive Attention**

Additive attention uses a feed-forward network which has a single hidden layer when computing attention value with query vector and key vector. Therefore, query vector and key vector do not have to have the same dimension, and the model is able to show consistent performance regardless of the size of the dimension. Additive attention used in [10] follows like:

$$a(\boldsymbol{Q},\boldsymbol{K}) = \omega_2^{\mathrm{T}} tanh(W_1[Q\;;K])$$

Concatenated query and key vector, and used a single hidden layer feed-forward network with tanh function as the activation function.

**Location-Aware Attention**

Location Aware Attention was proposed for speech recognition [12]. Unlike other attention mechanisms, this attention takes into account the distribution of the previous attention distribution. These distinctions make excellent for the model to be good at alignment. Thanks to these advantages, It has been a long time since it was proposed, but some ASR models are still using it.

**Multi-Head Attention**

Multi-head attention (MHA) was first explored in [26] for machine translation, and we extend this work to explore the value of MHA for speech.

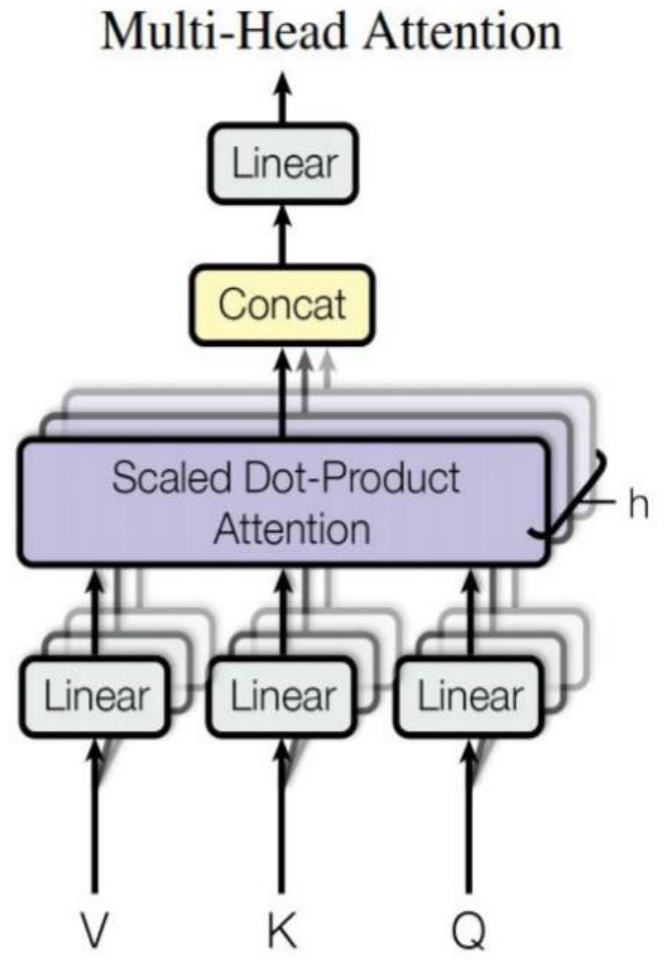

Figure 10: ***Multi-Head Attention consists of several attention layers running in parallel***.

Specifically, MHA extends the conventional attention mechanism to have multiple heads, where each head can generate a different attention distribution. We observed that each head has a different role on attending the encoder output, which contains higher features from inputs. We use 4 heads in the baseline model.

### 3.3.4 Optimization

We employ various method for optimization. In this section, we deal with three representative things: *Scheduled Sampling, Label Smoothing, Learning Rate Scheduling*. We observed that acoustic modeling is important, but the use of these techniques greatly influences the performance of the model.

**Scheduled Sampling**

Teacher forcing that feeding ground-truth label as the previous prediction helps the decoder to learn quickly at the beginning but causes a discrepancy between training and inference. The other method that samples from the probability distribution of the previous prediction and then uses the token to feed as the previous token when predicting the next label helps reduce the gap between training and inference behavior. Scheduled sampling process uses teacher forcing at the beginning of training steps, and in the training proceeds, linearly increases the probability of sampling from the model's prediction to 0.8 at the specific step, which then keeps constant until the end of training.

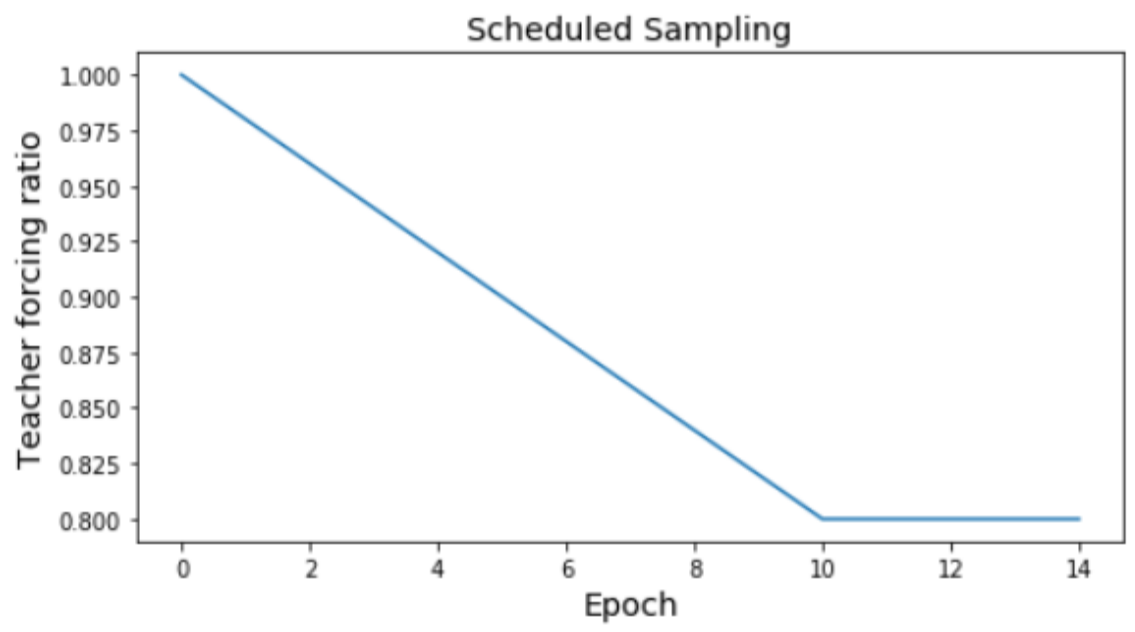

Figure 11: ***Teacher forcing scheduling*** *Reduced by 2% per epoch.*

The formula is as following:

$$\text{teacher forcing ratio} = max(1.0 - 0.02 \times epoch, 008)$$

Although other research said the impact of teacher forcing ratio on the result is slight but on the contrary, it was an important parameter for our experiment.

**Label Smoothing**

Label smoothing, which is proposed in [46] first, is frequently used in speech recognition tasks [13, 43]. Label smoothing helps to make the model less confident in its predictions and is a regularization mechanism that has successfully been applied in vision and speech task in [12, 48]. It encourages the model to have higher entropy at its prediction, and therefore makes the model more adaptable. We also followed the same design proposed in [13, 46] by smoothing the ground truth label distribution with a uniform distribution over all labels. We used 0.1 as an epsilon (smoothing).

**Learning Rate Scheduling**

The learning rate schedule is an important factor in determining the performance of ASR networks, especially so when augmentation is present. We warm-up learning rate for 400 steps from 0 to 3e-04. Afterwards, the learning rate is reduced whenever the loss on the validation set has not decreased by more than a certain threshold over a given number of epochs (default value: 1). This scheduling is controlled by PyTorch`s *ReduceLROnPlateau* method.

## 4. EXPERIMENTS

We conduct experiments on the KsponSpeech dataset which consists of 1,000 hours of labeled speech. We extract 80-dimensional log mel-filterbank coefficients features computed every 10ms with a 20ms window.

We use the Adam optimizer [48] and learning rate schedule with 400 warm-up steps and a peak learning rate of 0.0003. A weight decay with $10^{-6}$ weight is also added to all the trainable weights in the network.

We increased data by SpecAugment [43, 49] with mask parameter ($F = 20$), ten time masks with maximum time-mask ratio ($p_S = 0.05$), where the maximum size of the time mask is set to $p_S$ times the length of the utterance. The dataset provided from the AI Hub also has short utterances. Therefore, we followed the design proposed in [49] masking in proportion to the total length. We don't use Time warping.

Table 4: ***CER(%) of each feature with VGG Net extractor and multi-head attention.*** *The numbers in parentheses are feature size.*

| Feature | CER(%) |
| --- | --- |
| # MFCC (40) | 17.31 |
| # Log mel spectrogram (161) | 15.79 |
| # Log spectrogram (161) | 10.72 |
| **# Filter Bank (80)** | **10.31** |

Table 5: ***CER(%) of each CNN Extractor with filter bank and multi-head attention.***

| CNN Extractor | CER(%) |
| --- | --- |
| # Deep Speech 2 | 14.81 |
| **# VGG Net** | **10.31** |

Table 6: ***CER(%) of each Attention mechanism with CNN Extractor and filter bank.***

| Attention Mechanism | CER(%) |
| --- | --- |
| # Dot Product Attention | 18.33 |
| # Additive Attention | 16.87 |
| # Location Aware Attention | 13.52 |
| **# Multi-Head Attention** | **10.31** |

We use character error rate (CER) as a metric:

$$D = Distance_{LEV}(X, Y), CER(\%) = \frac{D}{L} \times 100$$

where X, Y are predicted and a ground truth script. The distance D is the Levenshtein distance between X, Y and the length L is a length of ground truth script Y. By this metric, we achieved 10.31% CER only using acoustic models. Also, we conducted many experiments with various options which KoSpeech provides. There are about 50 options but, in this section, we focused on the three parts: feature, CNN extractor, attention mechanism. The remaining conditions were all the same compared to the best recorded parameter, and see Table 4, 5, 6 for details.

Table 7: ***Prediction result when decoding process takes greedy search and beam search.*** *k means beam size.*

| | |
|---|---|
| **ORIGINAL** | 놀 만큼 논 거고 놀고 한 거지 |
| **GREEDY** | 놀만큼 농구가 또 놀고 한 거지 |
| **BEAM (k=3)** | 놀만큼 농구거 같애 놀고 |

Also, we employed beam search. Usually, beam search shows higher performance than greedy search and ensures an enriched expression of prediction. In our experiment, averagely, beam search decoding CER is higher 2% than greedy search.

## 5. CONCLUSION

Various features were experimented, and results depended on conditions, for example, RNN structure, ratio of listener layer and speller layer size, etc.

For performance gap between log spectrogram and filter bank, we guess that the implicit amount of information in log spectrogram was higher than filter bank but envelope information that expresses pronunciation was more explicit in filter bank. About MFCC, we interpret the degradation of performance as the effect of dimension reduction.

Not only features, but also the decoding mechanism showed interesting results. Greedy decoding mechanism performs better than beam search in our experiment. We think this is because of the absence of an external language model to rescore prediction. Internal language model is powerful for itself, but more sophisticated rescoring system still seems to be necessary during the beam search decoding step.

We planned multiple further works to improve *KoSpeech*. First is to shallowly fusion an external language model into model [45]. Second is to support more various recognition units, not only character but also grapheme, sub-words, are our next goal. Last, to add a transformer structure into *KoSpeech* toolkit to reduce learning time than before.

We introduce *KoSpeech*, open-source software for Korean automatic speech recognition (ASR) toolkit based on the deep learning library PyTorch. We hope to further develop *KoSpeech* to maintain strong speech recognition performance at the research frontier, providing a stable and expandable open-source toolkit.

## 6. ACKNOWLEDGEMENT

Thank you for your interest in the toolkit, even though it is still a lot lacking. Spoken Language Laboratory (of Sogang University) shared a lot of insights for us. Plus, we thank to Clova AI, Naver Corp. for making public ClovaCall [50]. At preprocessing transcript, Soyoung Cho and Jeongwon Kwak, of Kwangwoon University, helped us a lot.